\documentclass[letterpaper,11pt]{article}

\usepackage{times}

\usepackage{amsthm,amsmath,amsfonts}

\usepackage{graphicx}
\usepackage[small]{caption}
\usepackage[hidelinks]{hyperref}
\newcommand{\orcid}[1]{\,\href{https://orcid.org/#1}{\includegraphics[width=8pt]{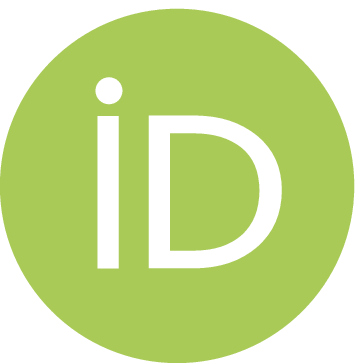}}}

\usepackage{color}

\usepackage{listings}

\title{On universal covers for carpenter's rule folding}

\author{Rain Jiang\orcid{0000-0002-0144-942X}\qquad
Kai Jiang\orcid{0000-0001-8165-0571}\qquad
Minghui Jiang\orcid{0000-0003-1843-9292}\,\thanks{\texttt{ dr.minghui.jiang at gmail.com}}\medskip\\
Home School, USA}

\date{}

\begin{document}

\maketitle

\begin{abstract}
	We present improved universal covers for carpenter's rule folding in the plane.
\end{abstract}

\section{Introduction}

A carpenter's rule is a concatenation of straight rulers of various lengths,
where consecutive rulers are joined by hinges
so that their angles can be adjusted independently.
Mathematically,
a carpenter's rule is distinguished
by the sequence of lengths of the individual rulers,
and a \emph{fold} of a carpenter's rule in the plane
is a polygonal chain of (possibly crossing) line segments with the given lengths.

A \emph{universal cover}, in the context of carpenter's rule folding,
is a planar set $C$ of diameter $1$ such that,
for any carpenter's rule $Z$ with maximum segment length $1$,
there exists a fold of $Z$ inside $C$.
We are interested in finding universal covers of the smallest possible area.

\begin{figure}[htb]
\centering\includegraphics{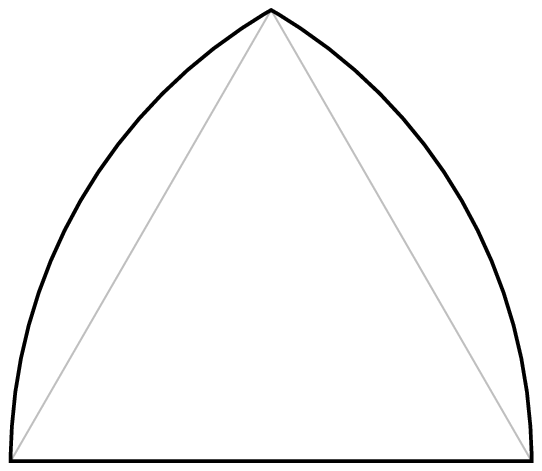}
\caption{A convex universal cover of area $\frac{\pi}3 - \frac{\sqrt3}4 = 0.614\ldots$.}
\label{fig:one_edge}
\end{figure}

C\u{a}linescu and Dumitrescu~\cite{CD05}
showed that the convex shape $R_2$ illustrated in Figure~\ref{fig:one_edge}
is a universal cover,
and hence established the first upper bound of $0.614\ldots$
on the minimum area of any universal cover for carpenter's rule folding.
From the other direction,
Klein and Lenz~\cite{KL07}
gave a lower bound of $0.475$ on the area of any universal cover
that is a convex set.

Refer to Figure~\ref{fig:one_edge}.
Start with an equilateral triangle $uvw$ of side length $1$,
with $u$ on the left, $v$ on the right, and $w$ at the top.
The shape $R_2$ is bounded
from the left by a circular arc $uw$ with circle center at $v$,
from the right by a circular arc $vw$ with circle center at $u$,
and from below by the segment $uv$.
Consider the set $(R_2)$ consisting of points on the two arcs $uw$ and $vw$,
including the three vertices $u$, $v$, and $w$.
For any point $p \in (R_2)$,
and for any length $l \in (0, 1]$,
there exists at least one other point $q \in (R_2)$ such that
the segment $pq$ has length exactly $l$ and is completely contained in $R_2$.
This property of $(R_2)$ allows a simple online strategy for folding any carpenter's rule
with maximum segment length $1$ inside $R_2$.

Recently,
Chen and Dumitrescu~\cite{CD15}
constructed a smaller, but nonconvex, universal cover,
thereby improving the upper bound on the minimum area of a universal cover
from $0.614\ldots$ to $0.583$.
They also gave a lower bound of $0.073$ on the area of any universal cover
that is simply connected but not necessarily convex.

Alt, Buchin, Cheong, Hurtado, Knauer, Schulz, and Whitesides~\cite{ABC+06}
studied $k$-universal covers for folding carpenter's rules with at most $k$ segments,
where $k$ is a small constant.

In this paper, we construct a sequence of even smaller, nonconvex universal covers,
and improve the upper bound on the minimum area of any universal cover to $0.55536036\ldots$.
All our constructions satisfy a similar property as $(R_2)$ for online folding.

\section{Two-edge cut}

\begin{figure}[htb]
\centering\includegraphics{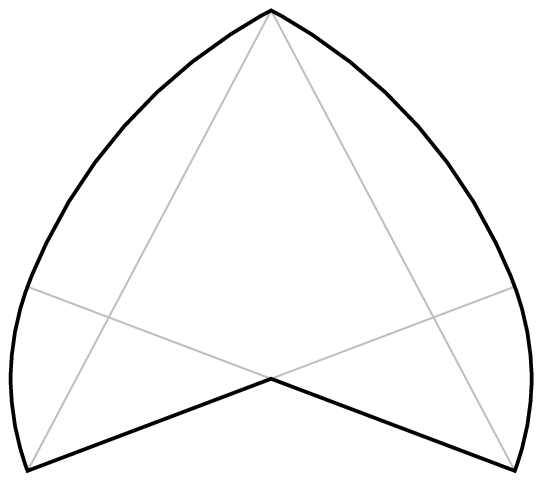}
\caption{A nonconvex universal cover of area $0.5726\ldots$ generated by a two-edge cut.}
\label{fig:two_edge}
\end{figure}

Refer to Figure~\ref{fig:two_edge}.
Let $a$ be an angle (in radians) to be determined.
Let $c$ be another angle (in radians) and $x_0$ be a length, that depend on $a$.
Our first construction consists of four overlapping circular sectors:
two large sectors of radius $1$ and angle $a$,
and two opposing small sectors of radius $\frac12$ and angle $2c$.
The union of the four sectors is a nonconvex shape
with a two-edge cut at the bottom.

Consider the overlapped region of the shape, which is the difference of two isosceles triangles
sharing	the same base:
the large triangle has two equal sides $1$ and two equal angles $a + c$;
the small triangle (resulting from the two-edge cut)
has two equal sides $\frac12$ and two equal angles $c$.
The length $x_0$ of the shared base can be expressed in two different ways:
\begin{equation}\label{eq:two_edge}
x_0 = 2 \cos(a + c) = \cos c.
\end{equation}
Take derivative of these two expressions of $x_0$ with respect to $c$, and we get
\begin{equation}\label{eq:two_edge_d}
-2 \sin(a + c) \left( \frac{da}{dc} + 1 \right)
= -\sin c.
\end{equation}

The area of the overlapped region is
\begin{align*}
\frac{x_0}2 \left( \sin (a + c) - \frac12 \sin c \right)
&=
\cos(a + c) \sin(a + c)
- \frac14 \cos c \sin c\\
&=
\frac12 \sin (2a + 2c) - \frac18 \sin 2c.
\end{align*}
The area $A$ of the shape is the total area of the four sectors
minus the overlapped area:
$$
A = a + \frac c2 - \frac12 \sin (2a + 2c) + \frac18 \sin 2c.
$$
Then,
\begin{align*}
\frac{dA}{dc}
&=
\frac{da}{dc}
+ \frac12 - \frac12 \cos(2a + 2c) \left( 2 \frac{da}{dc} + 2 \right)
+ \frac14 \cos 2c\\
&=
\left( \frac{da}{dc} + 1 \right)
- \frac12 - \cos(2a + 2c) \left( \frac{da}{dc} + 1 \right)
+ \frac14 \cos 2c\\
&=
\left( \frac{da}{dc} + 1 \right)
(1 - \cos(2a + 2c))
+ \frac14 (1 + \cos 2c)
- \frac34\\
&=
\left( \frac{da}{dc} + 1 \right)
2 \sin\left( a + c \right)
\sin\left( a + c \right)
+ \frac12 \cos c
\cos c
- \frac34\\
&=
\sin c \sin(a + c)
+ \cos(a + c) \cos c
- \frac34,
\end{align*}
where the last step uses \eqref{eq:two_edge_d} and \eqref{eq:two_edge}.
It follows that
$$
\frac{dA}{dc}
=
\cos(a + c - c)
- \frac34
=
\cos a
- \frac34.
$$
Let $a = \arccos\frac34 \approx 41.4^\circ$ so that $\frac{dA}{dc} = 0$.
Then $c$ and $x_0$ are determined by~\eqref{eq:two_edge}.
Incidentally, $a = 2c$.
Finally,
$$
A = \frac54 a - \frac12 \sin 3a + \frac18 \sin a = 0.5726\ldots.
$$

\section{Three-edge cut}

\begin{figure}[htb]
\centering\includegraphics{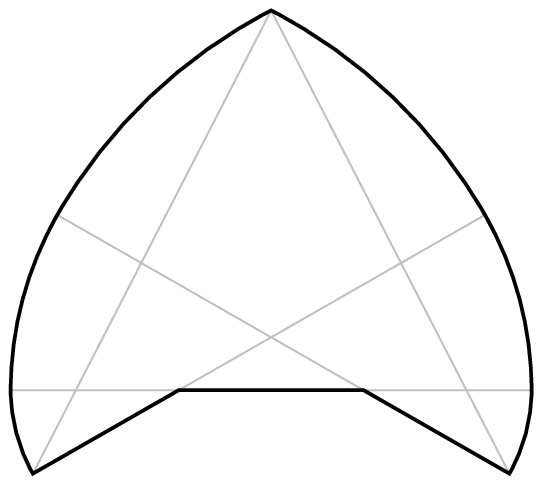}
\caption{A nonconvex universal cover of area $0.5635\ldots$ generated by a three-edge cut.}
\label{fig:three_edge}
\end{figure}

Refer to Figure~\ref{fig:three_edge}.
Let $a$ and $b$ be two angles (in radians) to be determined.
Let $x_0$, $x_1$, and $x_2$, where $x_1 + x_2 + x_1 = 1$,
be three lengths that depend on $a$ and $b$.
Our construction in this section consists of six overlapping circular sectors:
two small sectors of radius $x_1$,
two medium sectors of radius $x_1 + x_2 = 1 - x_1$,
and two large sectors of radius $x_1 + x_2 + x_1 = 1$.
The two large sectors have the same angle $a$;
the other four sectors have the same angle $b$.
Moreover, each medium sector is directly opposite to a small sector
so that the sum of their radii is exactly $1$.
The union of the six sectors is a nonconvex shape with a three-edge cut at the bottom.

In clockwise direction, the left boundary of the shape consists of circular arcs
from three of the six sectors, one small, one medium, and one large.
These three sectors are adjacent but interior-disjoint.
They have radii $x_1$, $1 - x_1$, $1$ and angles $b$, $b$, $a$ respectively,
so their total area is
$$
\frac b2 \left( x_1^2 + (1 - x_1)^2 \right) + \frac a2.
$$
The situation is symmetric for the right boundary of the shape.
The area of the shape is the total area of the six sectors
minus the area of the overlapped region in the middle.

The overlapped region of the shape is the difference
of an isosceles triangle and an isosceles trapezoid:
the triangle has equal sides $1$ and base $x_0$;
the trapezoid (resulting from the three-edge cut)
has equal sides $x_1$, upper base $x_2$, and lower base $x_0$.
The angle $a+b$ of the triangle and the angle $b$ of the trapezoid require that
$$
\cos(a+b) = \frac{x_0}2 \Big/ 1
\qquad
\textrm{and}
\qquad
\cos b = \frac{x_0 - x_2}2 \Big/ x_1.
$$
In addition, we require that $x_1 + x_2 + x_1 = 1$.
These three constraints together determine the three lengths $x_0$, $x_1$, and $x_2$ uniquely:
$$
x_0 = 2\cos(a+b),
\qquad
x_1 = \frac{1 - x_0}{2(1 - \cos b)},
\qquad
x_2 = 1 - 2 x_1.
$$
Then the overlapped area is
$$
\frac{x_0}2 \sin(a+b)
- \frac{x_0 + x_2}2 x_1\sin b.
$$
So the area of the shape is
$$
A = 
b \left( x_1^2 + (1 - x_1)^2 \right) + a
- \frac{x_0}2 \sin(a+b)
+ \frac{x_0 + x_2}2 x_1\sin b.
$$

A calculation shows that
$A = 0.5635\ldots$
when $a = 0.575939 \approx 33.0^\circ$
and $b = 0.519805 \approx 29.8^\circ$.
Note that the construction
that gives the previous best upper bound of $0.583$~\cite{CD15}
is a special case of our construction with the angle $a$ set to $0$.

\section{Four-edge cut}

\begin{figure}[htb]
\centering\includegraphics{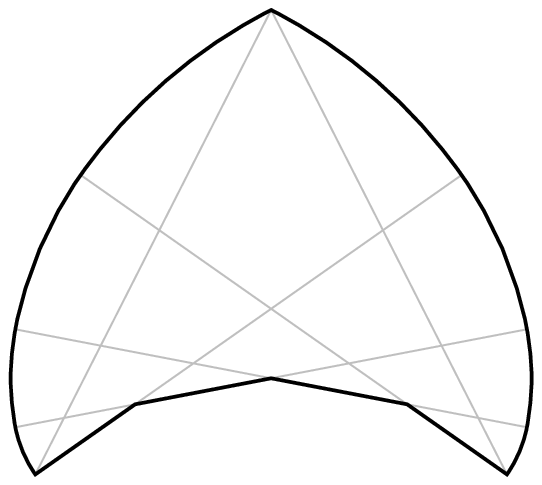}
\caption{A nonconvex universal cover of area $0.5600\ldots$ generated by a four-edge cut.}
\label{fig:four_edge}
\end{figure}

Refer to Figure~\ref{fig:four_edge}.
Let $a$, $b$, and $c$ be three angles (in radians) to be determined.
Let $x_0$, $x_1$, $x_2$, and $x_3$, where $x_1 + x_3 = \frac12$,
be four lengths that depend on $a$, $b$, and $c$.
Our construction in this section consists of eight overlapping circular sectors:
two tiny sectors of radius $x_1$,
two small sectors of radius $x_1 + x_3 = \frac12$,
two medium sectors of radius $x_1 + x_3 + x_3 = 1 - x_1$,
and two large sectors of radius $x_1 + x_3 + x_3 + x_1 = 1$.
The two large sectors have the same angle $a$;
the two medium sectors and the two tiny sectors have the same angle $b$;
the two small sectors have the same angle $2c$.
Moreover,
each medium sector is opposite to a tiny sector,
and the two small sectors are opposite to each other,
so that the sum of radii of each pair of opposing sectors is $1$.
The union of the eight sectors is a nonconvex shape with a four-edge cut at the bottom.

In clockwise direction, the left boundary of the shape consists of circular arcs
from four of the eight sectors, one tiny, one small, one medium, and one large.
These four sectors are adjacent but interior-disjoint.
They have radii $x_1$, $\frac12$, $1 - x_1$, $1$ and angles $b$, $2c$, $b$, $a$ respectively,
so their total area is
$$
\frac b2 \left( x_1^2 + (1 - x_1)^2 \right) + \frac c4 + \frac a2.
$$
The situation is symmetric for the right boundary of the shape.
The area of the shape is the total area of the eight sectors
minus the area of the overlapped region in the middle.

The overlapped region of the shape is the difference of a triangle and a pentagon.
The triangle is isosceles with equal sides $1$ and base $x_0$.
The pentagon (resulting from the four-edge cut)
is the union of
an isosceles trapezoid (with equal sides $x_1$, upper base $x_2$, and lower base $x_0$)
and a smaller isosceles triangle (with equal sides $x_3$ and base $x_2$).
The angle $a + b + c$ of the triangle with equal sides $1$,
the angle $b + c$ of the trapezoid,
and the angle $c$ of the triangle with equal sides $x_1$
require that
$$
\cos(a + b + c) = \frac{x_0}2 \Big/ 1,
\qquad
\cos(b + c) = \frac{x_0 - x_2}2 \Big/ x_1,
\qquad
\cos c = \frac{x_2}2 \Big/ x_3.
$$
In addition, we require that $x_1 + x_3 = \frac12$.
These four constraints together determine the four lengths $x_0$, $x_1$, $x_2$, and $x_3$ uniquely:
$$
x_0 = 2\cos(a + b + c),
\qquad
x_1 = \frac{\cos c - x_0}{2\cos c - 2\cos(b + c)},
\qquad
x_2 = (1 - 2 x_1)\cos c,
\qquad
x_3 = \frac12 - x_1.
$$
Then the overlapped area is
$$
\frac{x_0}2 \sin(a + b + c)
- \frac{x_0 + x_2}2 x_1\sin(b + c)
- \frac{x_2}2 x_3 \sin c.
$$
So the area of the shape is
$$
A = b \left( x_1^2 + (1 - x_1)^2 \right) + \frac c2 + a
- \frac{x_0}2 \sin(a + b + c)
+ \frac{x_0 + x_2}2 x_1\sin(b + c)
+ \frac{x_2}2 x_3 \sin c.
$$

A calculation shows that $A = 0.5600\ldots$
when $a = 0.488669 \approx 28.0^\circ$,
$b = 0.423144 \approx 24.2^\circ$,
and $c = 0.189158 \approx 10.8^\circ$.

\section{Involute method and smooth cut}

The universal covers described in the previous three sections are all constructed by the same method,
which we call the \emph{involute method}.
An \emph{involute} of a curve is the locus of a point on a piece of taut string as the string is
either unwrapped from or wrapped around the curve.

The involute method works as follows.
Start with a pseudotriangle $uvw$ consisting of a segment $uw$, a segment $vw$, and a curve $uv$,
all of the same length $1$.
Place $uvw$ such that the segment $uv$ is horizontal,
with $u$ on the left, $v$ on the right, and $w$ above them.
Imagine a string of length $1$ wrapped around the curve $uv$ from above.
As we unwrap the string clockwise with its right end fixed at $v$,
its left end traces an arc from $u$ to $w$.
Symmetrically,
as we unwrap the string counterclockwise with its left end fixed at $u$,
its right end traces an arc from $v$ to $w$.
Then the resulting shape,
bounded from the left by the arc $uw$,
from the right by the arc $vw$,
and from below by the arc $uv$,
is a universal cover,
because the points on the two arcs $uw$ and $vw$ satisfy a similar property as $(R_2)$ mentioned
earlier that enables an online folding strategy.

When the curve $uv$ is just a single line segment,
the resulting shape is exactly the convex shape $R_2$ illustrated in Figure~\ref{fig:one_edge},
which holds the current record of $0.614\ldots$
for the area of a convex universal cover.
The curves $uv$ in our constructions
illustrated in Figures~\ref{fig:two_edge}, \ref{fig:three_edge}, and~\ref{fig:four_edge}
are polygonal chains consisting of two, three, and four edges, respectively.
By the involute method,
they generate nonconvex universal covers of areas
$0.5726\ldots$, $0.5635\ldots$, and $0.5600\ldots$.

\begin{figure}[htb]
	\centering\resizebox{\linewidth}{!}{\includegraphics{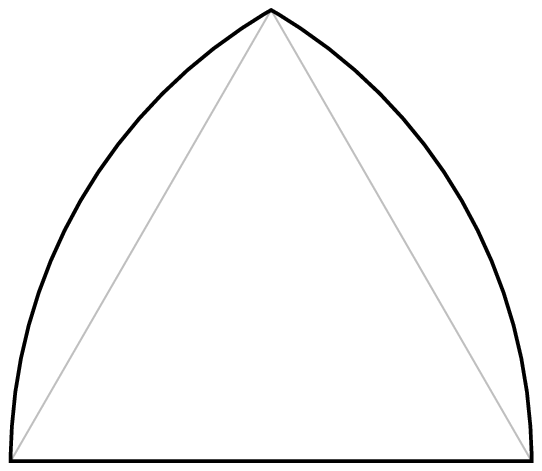}\includegraphics{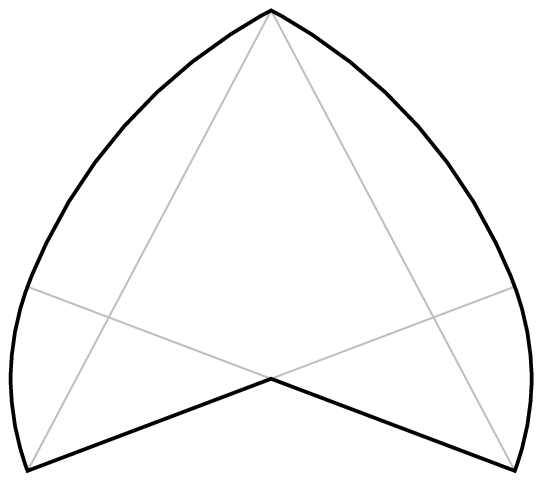}\includegraphics{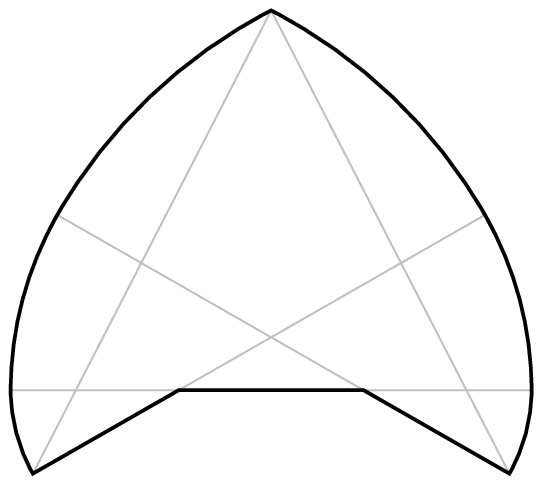}\includegraphics{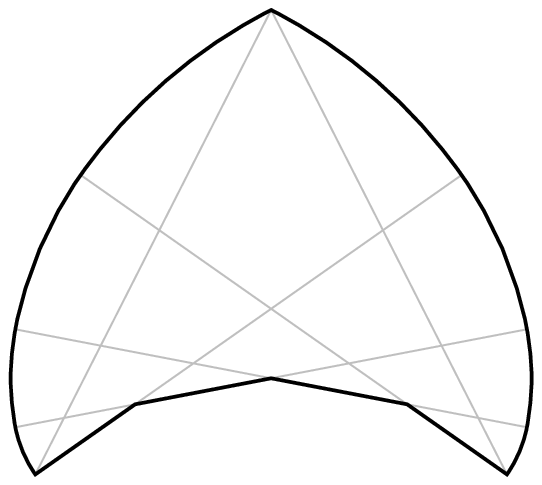}}
	\caption{A sequence of universal covers constructed by the involute method.}
\label{fig:involute}
\end{figure}

Refer to Figure~\ref{fig:involute}.
Following the trend that more edges lead to smaller area,
we wrote a computer program that uses a local search heuristic
to generate universal covers from
polygonal chains of many edges.
As the number of edges increases to the thousands,
the minimum area of universal covers found by the computer program
steadily decreases and seems to converge to $0.5553603\ldots$.
In the following,
we construct a smooth curve $uv$,
which generates a nonconvex universal cover of area $0.55536036\ldots$.

For any two points $p$ and $q$,
we denote by $pq$
either a line segment or some curve (for example, a circular arc)
with $p$ and $q$ as endpoints,
when the exact meaning is clear from context.
Also, we denote by $|pq|$ the distance between two points $p$ and $q$,
and denote by $|\vec v|$ the length (or magnitude, norm) of a vector $\vec v$.

\begin{figure}[htb]
\centering\includegraphics{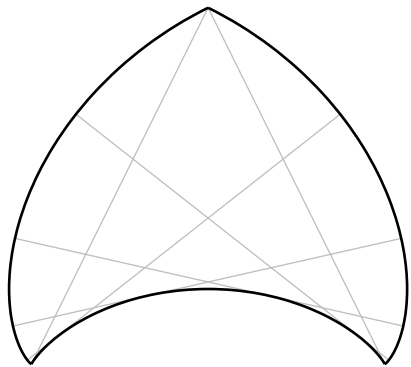}
\caption{A nonconvex universal cover of area $0.55536036\ldots$ generated by a smooth cut.}
\label{fig:smooth}
\end{figure}

Refer to Figure~\ref{fig:smooth}.
Let $a$ be an angle (in radians) to be determined.
We will construct a smooth, concave arc $uv$ through the origin $o$ and symmetric about the $y$-axis,
with $u$ on the left and $v$ on the right.
Let $w$ be the point above the arc $uv$ such that $|uw| = |vw| = 1$.
We will ensure that the arc $uv$ has three properties.
First, its length is exactly $1$.
Second, it meets the two segments $uw$ and $vw$ at zero angles at $u$ and $v$, respectively.
Third, its radius of curvature is zero at $u$ and $v$.
The two segments $uw$ and $vw$, together with the arc $uv$,
form a pseudotriangle.

Unwrap the arc $uv$ clockwise from $u$ to $v$, and we get an involute of $uv$
which is an arc between $u$ and $w$.
Symmetrically, unwrap the arc $uv$ counterclockwise from $v$ to $u$, and we get another involute
of $uv$ which is an arc between $v$ and $w$.
Thus we obtain a nonconvex shape bounded
from the left by the arc $uw$,
from the right by the arc $vw$,
and from below by the arc $uv$.

\paragraph{Deriving equations for the arc $uv$}

Let $\vec c_0(t) = (x_0(t), y_0(t))$ for $t \in [-a, a]$
be the parametric representation of the arc $uv$,
where the three points $u$, $o$, and $v$
correspond to $\vec c_0(-a)$, $\vec c_0(0)$, and $\vec c_0(a)$, respectively,
and the parameter $t$ is the clockwise angle from the $y$ axis to the normal of the arc $uv$
at $\vec c_0(t)$.

Consider $\vec c_0\,\!\!'(t) = (x_0'(t), y_0'(t))$,
the derivative of
$\vec c_0(t) = (x_0(t), y_0(t))$
with respect to the parameter $t$.
Define the function $g(t) = b_0 + b_1 \cos t + b_2 \cos 2t$,
where the three coefficients $b_0$, $b_1$, and $b_2$
are real numbers depending on $a$.
We design the vector $\vec c_0\,\!\!'(t)$ to have length
$|\vec c_0\,\!\!'(t)| = g(t)$
and direction
$\frac{\vec c_0\,\!\!'(t)}{|\vec c_0\,\!\!'(t)|} = (\cos t, -\sin t)$
by letting
$x_0'(t) = g(t)\cos t$,
$y_0'(t) = -g(t)\sin t$.
Then,
\begin{align*}
	x_0'(t) &= \phantom{-}(b_0 + b_1\cos t + b_2\cos 2t)\cos t,\\
	y_0'(t) &= -(b_0 + b_1\cos t + b_2\cos 2t)\sin t,\\
	\int x_0'(t) \,dt &= b_0\sin t
	+ b_1\frac14(2t + \sin 2t)
	+ b_2\frac16(\sin 3t + 3\sin t),\\
	\int y_0'(t) \,dt &= b_0\cos t
	+ b_1\frac14\cos 2t
	+ b_2\frac16(\cos 3t - 3\cos t).
\end{align*}

To ensure that the arc $uv$ goes through the origin $o$,
we require that $x_0(0) = 0$ and $y_0(0) = 0$. Thus,
\begin{align*}
	x_0(t) &= \frac1{12}\big(6b_1 t + 6(2b_0 + b_2)\sin t + 3b_1\sin 2t + 2b_2\sin 3t\big),\\
	y_0(t) &= \frac1{12}\big(6(2b_0 - b_2)\cos t + 3b_1\cos 2t + 2b_2\cos 3t - 12b_0 - 3b_1 + 4b_2\big).
\end{align*}
Since $x_0(-t) = -x_0(t)$ and $y_0(-t) = y_0(t)$ for $t \in [-a, a]$,
the arc $uv$ is symmetric about the $y$-axis.

\paragraph{Deriving equations for the arc $uw$}

Let $\vec c_1(t) = (x_1(t), y_1(t))$ for $t \in [-a, a]$ be the parametric representation
of the arc $uw$ which is the involute of the arc $uv$ as we unwrap it clockwise from $u$ to $v$.
By the definition of involute, we have,
$$
\vec c_1(t)
= \vec c_0(t)
- \frac{ \vec c_0\,\!\!'(t) }{| \vec c_0\,\!\!'(t) |}
\ell(t),
$$
where
$$
\ell(t) =
\int_{-a}^t |\vec c_0\,\!\!'(s)| \,ds
$$
is the length of the subarc of $uv$ with parameter from $-a$ to $t$.

Recall that
$\frac{\vec c_0\,\!\!'(t)}{|\vec c_0\,\!\!'(t)|} = (\cos t, -\sin t)$
and
$|\vec c_0\,\!\!'(t)| = g(t)$.
The indefinite integral
$\int |\vec c_0\,\!\!'(t)| \,dt$
evaluates to
$$
h(t)
= \int g(t) \,dt
= \int (b_0 + b_1\cos t + b_2\cos 2t) \,dt
= b_0 t + b_1\sin t + \frac12 b_2\sin 2t,
$$
and it follows that
$$
\ell(t)
= \int_{-a}^t g(t) \,ds
= h(t) - h(-a)
= h(t) + h(a).
$$
Therefore we have
\begin{align*}
	x_1(t) &= x_0(t) - \ell(t) \cos t,\\
	y_1(t) &= y_0(t) + \ell(t) \sin t.
\end{align*}

\paragraph{Deriving equations for the arc $vw$}

The arc $vw$ is the involute of the arc $uv$ as we unwrap it counterclockwise from $v$ to $u$.
Although the two arcs $uw$ and $vw$ are symmetric about the $y$ axis, and are both
derived by unwrapping the arc $uv$,
the unwrapping for $vw$ happens in opposite direction as the unwrapping for $uw$.
We could of course change the parameter $t$ to mean the counterclockwise angle,
then derive new equations for the arc $uv$,
and then derive equations for the arc $vw$ accordingly.
But there is an easier way.

Let $\vec c_2(t) = (x_2(t), y_2(t))$ for $t \in [-a, a]$ be the parametric representation
of the arc $vw$, where the parameter $t$ means clockwise angle as before.
Reverse the unwrapping process for $vw$ from $t = a$ to $-a$,
and we get a wrapping process from $t = -a$ to $a$,
where a taut string of length $1$ begins as the segment $uw$ at $t = -a$,
and ends as the arc $uv$ at $t = a$.

Fix any $t \in [-a, a]$.
The vector $\vec c_1(t) - \vec c_0(t)$ points to the direction
$(-\cos t, \sin t)$,
and thus the vector $\vec c_2(t) - \vec c_0(t)$ points to the opposite direction
$(\cos t, -\sin t)$,
both perpendicular to the normal of the arc $uv$ at $\vec c_0(t)$.
The two vectors 
$\vec c_1(t) - \vec c_0(t)$
and $\vec c_2(t) - \vec c_0(t)$
correspond to two complementary parts of the arc $uv$,
from $-a$ to $t$,
and from $t$ to $a$, respectively.
Together, their total length is exactly the length of the arc $uv$, which is $1$.
Thus $\vec c_2(t) - \vec c_1(t) = (\cos t, -\sin t)$,
and we have
\begin{align*}
	x_2(t) &= x_1(t) + \cos t,\\
	y_2(t) &= y_1(t) - \sin t.
\end{align*}

\paragraph{Determining the coefficients}

The length of the arc $uv$ can be expressed as
$$
\int_{-a}^a |\vec c_0\,\!\!'(t)| \,dt
= \int_{-a}^a g(t) \,dt
= h(a) - h(-a)
= 2h(a).
$$
To ensure that this length is equal to $1$,
we require that $2h(a) = 1$. Thus,
\begin{equation}\label{eq:length1}
	2b_0 a + 2b_1\sin a + b_2\sin 2a = 1.
\end{equation}

Recall that $\frac{\vec c_0\,\!\!'(t)}{|\vec c_0\,\!\!'(t)|} = (\cos t, -\sin t)$.
To ensure that the arc $uv$ meets the two segments $uw$ and $vw$ at zero angles,
we require that $x_0(a) = \cos a$. Thus,
\begin{equation}\label{eq:tangent}
	6b_1 a + 6(2b_0 + b_2)\sin a + 3b_1\sin 2a + 2b_2\sin 3a = 12\cos a.
\end{equation}

Recall that $|\vec c_0\,\!\!'(t)| = g(t)$,
where $g(t) = b_0 + b_1 \cos t + b_2 \cos 2t$.
To ensure that the arc $uv$ has zero radius of curvature
at the two endpoints $u$ and $v$,
we require that $g(-a) = g(a) = 0$. Thus,
\begin{equation}\label{eq:b0b1b2}
	b_0 + b_1\cos a + b_2\cos 2a = 0.
\end{equation}

Together,
the three equations \eqref{eq:length1}, \eqref{eq:tangent}, and~\eqref{eq:b0b1b2}
are sufficient to determine the three coefficients $b_0$, $b_0$, and $b_1$
as functions of $a$.
From~\eqref{eq:b0b1b2},
we get
\begin{equation}\label{eq:b0}
b_0 = -b_1\cos a - b_2\cos 2a.
\end{equation}
Substitute \eqref{eq:b0} in~\eqref{eq:length1}, and we get
\begin{gather*}
-2b_1 a\cos a
-2b_2 a\cos2a
	+ 2b_1\sin a + b_2\sin 2a = 1\\
	2b_1(\sin a - a\cos a) + b_2(\sin 2a - 2a\cos 2a) = 1.
\end{gather*}
Thus,
\begin{equation}\label{eq:b1}
b_1 = \frac{1 - b_2(\sin 2a - 2a\cos 2a)}{2(\sin a - a\cos a)}.
\end{equation}
Substitute \eqref{eq:b0} in~\eqref{eq:tangent}, and we get
\begin{gather*}
	6b_1 a
	- 12b_1\sin a\cos a
	- 12b_2\sin a\cos 2a
	+ 6b_2\sin a + 3b_1\sin 2a + 2b_2\sin 3a = 12\cos a\\
	3b_1(2a - 4\sin a\cos a + \sin 2a) + 2b_2(-6\sin a\cos 2a + 3\sin a + \sin 3a) = 12\cos a\\
	3b_1(2a - \sin 2a) + 2b_2(8\sin^3 a) = 12\cos a\\
	3b_1(2a - \sin 2a) = 12\cos a - 16b_2\sin^3 a.
\end{gather*}
Substitute \eqref{eq:b1} in the above equation, and it follows that
$$
3(1 - b_2(\sin 2a - 2a\cos 2a))(2a - \sin 2a) = 2(\sin a - a\cos a) (12\cos a - 16b_2\sin^3 a).
$$
Thus,
\begin{align}\label{eq:b2}
	b_2
	&= \frac{
	2(\sin a - a\cos a)12\cos a - 3(2a - \sin 2a)
	}{
		2(\sin a - a\cos a)16\sin^3 a - 3(\sin 2a - 2a\cos 2a)(2a - \sin 2a)
	}\nonumber\\
	&= \frac{
		15\sin 2a
		- 24a\cos^2 a
		- 6a
	}{
		12a^2\cos 2a - 2a(16\sin^3 a\cos a + 3\sin 2a + 3\sin 2a\cos 2a)
		+ 32\sin^4 a + 3\sin^2 2a
	}\nonumber\\
	&= \frac{
		15\sin 2a
		- 6a(4\cos^2 a + 1)
	}{
		12a^2\cos 2a - 2a(8\sin 2a\sin^2 a + 6\sin 2a\cos^2 a)
		+ 2\sin^2 a(16\sin^2 a + 6\cos^2 a)
	}\nonumber\\
	&= \frac{
		15\sin 2a
		- 6a(2\cos 2a + 3)
	}{
		12a^2\cos 2a - 2a\sin 2a(8\sin^2 a + 6\cos^2 a)
		+ (1 - \cos 2a)(11 - 5\cos 2a)
	}\nonumber\\
	&= \frac{
		15\sin 2a
		- 6a(2\cos 2a + 3)
	}{
		12a^2\cos 2a - 2a\sin 2a(7 - \cos 2a)
		+ (1 - \cos 2a)(11 - 5\cos 2a)
	}.
\end{align}
By \eqref{eq:b2}, \eqref{eq:b1}, \eqref{eq:b0}, we can evaluate $b_2$, $b_1$, $b_0$
from $a$.

\paragraph{Calculating the area}

Recall that
$$
\ell(t)
= h(t) + h(a)
= b_0 t + b_1\sin t + \frac12 b_2\sin 2t + h(a).
$$
Write $b = h(a)$.
Then,
\begin{align}
	\int \ell(t)^2 \,dt
	&= \int \Big(b_0 t + b_1\sin t + \frac12 b_2\sin 2t + b\Big)^2 \,dt\nonumber\\
	&= \frac1{96}\Big( 32b_0^2 t^3 + 96b_0b t^2 + 12(4b_1^2 + b_2^2 + 8b^2) t\nonumber\\
	&\qquad + 48b_1(4b_0 + b_2)\sin t + 24(b_0b_2 - b_1^2)\sin 2t - 16b_1b_2\sin 3t - 3b_2^2\sin 4t\nonumber\\
	&\qquad - 48(b_0 t + b)(4b_1\cos t + b_2\cos 2t)
	\Big).\nonumber\\
	\int_{-a}^a \ell(t)^2 \,dt
	&= \frac1{48}\Big( 32b_0^2 a^3 + 12(4b_1^2 + b_2^2 + 8b^2) a\nonumber\\
	&\qquad + 48b_1(4b_0 + b_2)\sin a + 24(b_0b_2 - b_1^2)\sin 2a - 16b_1b_2\sin 3a - 3b_2^2\sin 4a\nonumber\\
	&\qquad - 48b_0 a(4b_1\cos a + b_2\cos 2a)
	\Big).
\end{align}

By analogy of summing up sector areas as we did in previous sections,
the integral above accounts for the total area of our nonconvex shape
except that an overlapped region is included twice.
This overlapped region is the difference of the triangle $uvw$ and the cap region
between the arc $uv$ and the segment $uv$.

The area of the triangle $uvw$ is simply
\begin{equation}
A_{uvw} = \cos a\sin a.
\end{equation}

The area of the cap region is
$$
A_{uv} = -2\int_0^a x_0(t) y_0'(t) \,dt.
$$
Recall the parametric equations for $x_0(t)$ and $y_0'(t)$.
Then,
\begin{align*}
	&
	288 \cdot (-2)\int x_0(t) y_0'(t) \,dt =
	48 \cdot (-12)\int x_0(t) y_0'(t) \,dt\\
	&\quad=
	48\int (6b_1 t + 6(2b_0 + b_2)\sin t + 3b_1\sin 2t + 2b_2\sin 3t)
	(b_0 + b_1\cos t + b_2\cos 2t)\sin t \,dt\\
	&\quad=
	12(24b_0^2 - 4b_2^2 + 3b_1^2)t
	+ 24b_1(21b_0 - 2b_2)\sin t
	- 12(12b_0^2 - 8b_0b_2 - 5b_2^2 - 3b_1^2)\sin 2t\\
	&\quad\quad
	- 4b_1(18b_0 - b_2)\sin 3t
	- 3(16b_0b_2 + 4b_2^2 + 3b_1^2)\sin 4t
	- 12b_1b_2\sin 5t
	- 4b_2^2\sin 6t\\
	&\quad\quad
	- 24b_1 t( 6(2b_0 - b_2)\cos t + 3b_1\cos 2t + 2b_2\cos 3t ),
\end{align*}
which evaluates to $0$ when $t = 0$.
Then,
\begin{align}
	A_{uv}
	&= 
	-2\int_0^a x_0(t) y_0'(t) \,dt\nonumber\\
	&=
	\frac1{288}\Big(
	12(24b_0^2 - 4b_2^2 + 3b_1^2)a
	+ 24b_1(21b_0 - 2b_2)\sin a
	- 12(12b_0^2 - 8b_0b_2 - 5b_2^2 - 3b_1^2)\sin 2a\nonumber\\
	&\qquad
	- 4b_1(18b_0 - b_2)\sin 3a
	- 3(16b_0b_2 + 4b_2^2 + 3b_1^2)\sin 4a
	- 12b_1b_2\sin 5a
	- 4b_2^2\sin 6a\nonumber\\
	&\qquad
	- 24b_1 a( 6(2b_0 - b_2)\cos a + 3b_1\cos 2a + 2b_2\cos 3a )
	\Big).
\end{align}

In summary, the total area of the nonconvex shape is
$$
A = \int_{-a}^a \ell(t)^2 \,dt - A_{uvw} + A_{uv}.
$$
A calculation using the GNU arbitrary precision calculator \texttt{bc}
(see appendix for source code and output)
shows that when
$$
a = 1.11073213677147211458454234766 \approx 63.64^\circ,
$$
and correspondingly,
\begin{align*}
	b_0 &=           -0.31003908380107665108233928\ldots\\
	b_1 &= \phantom{-}0.88242010074246605497268495\ldots\\
	b_2 &= \phantom{-}0.13498096758065222221003550\ldots,
\end{align*}
the area $A$ is $0.55536036\ldots$.

\section{Conclusion}

We conjecture that the smallest universal cover for carpenter's rule folding
can be generated by the involute method, and hence the task of finding the best universal cover
reduces to finding the best generating curve $uv$.
We chose the function $g(t) = b_0 + b_1 \cos t + b_2 \cos 2t$
in the previous section
because it gives a good approximation of the radius of curvature of the best curve
found by our computer program based on local search.
Following the general idea of Fourier analysis,
we could try to improve the upper bound further
by including more terms in $g(t)$ (the next one would be $b_3\cos 3t$),
but such efforts would likely be rewarded with diminishing returns.

In contrast to our upper bound of $0.55536036\ldots$,
the current best lower bound
is $0.475$ for convex universal covers~\cite{KL07},
and is $0.073$
for universal covers that are not necessarily convex but still simply connected~\cite{CD15}.
For universal covers that are not simply connected,
can we prove a constant lower bound above zero?

\appendix

\newpage
\section*{Source code \texttt{ smooth.bc}}
\lstset{basicstyle=\footnotesize\ttfamily, keywordstyle=\ttfamily,
	showstringspaces=false, tabsize=2 }
\begin{lstlisting}
scale = 60

define solve(a) {
	cosa  = c(a)
	cos2a = c(2 * a)
	cos3a = c(3 * a)

	sina = s(a)
	sin2a = s(2 * a)
	sin3a = s(3 * a)
	sin4a = s(4 * a)
	sin5a = s(5 * a)
	sin6a = s(6 * a)

	b2p = 15 * sin2a - 6 * a * (2 * cos2a + 3)
	b2q = 12 * a * a * cos2a - 2 * a * sin2a * (7 - cos2a) + (1 - cos2a) * (11 - 5 * cos2a)
	b2 = b2p / b2q

	b1 = (1 - b2 * (sin2a - 2 * a * cos2a)) / (2 * (sina - a * cosa))
	b0 = -(b1 * cosa + b2 * cos2a)
	b = b0 * a + b1 * sina + b2 * sin2a / 2

	print "\n"
	print "a  ", a, "\n"
	print "b0 ", b0, "\n"
	print "b1  ", b1, "\n"
	print "b2  ", b2, "\n"

	sec1 = 32 * b0 * b0 * a * a * a
	sec2 = 12 * (4 * b1 * b1 + b2 * b2 + 8 * b * b) * a
	sec3 = 48 * b1 * (4 * b0 + b2) * sina
	sec4 = 24 * (b0 * b2 - b1 * b1) * sin2a
	sec5 = -16 * b1 * b2 * sin3a
	sec6 = -3 * b2 * b2 * sin4a
	sec7 = -48 * b0 * a * (4 * b1 * cosa + b2 * cos2a)
	sectors = (sec1 + sec2 + sec3 + sec4 + sec5 + sec6 + sec7) / 48

	triangle = cosa * sina

	cap1 = 12 * (24 * b0 * b0 - 4 * b2 * b2 + 3 * b1 * b1) * a
	cap2 = 24 * b1 * (21 * b0 - 2 * b2) * sina
	cap3 = -12 * (12 * b0 * b0 - 8 * b0 * b2 - 5 * b2 * b2 - 3 * b1 * b1) * sin2a
	cap4 = -4 * b1 * (18 * b0 - b2) * sin3a
	cap5 = -3 * (16 * b0 * b2 + 4 * b2 * b2 + 3 * b1 * b1) * sin4a
	cap6 = -12 * b1 * b2 * sin5a
	cap7 = -4 * b2 * b2 * sin6a
	cap8 = -24 * b1 * a * (6 * (2 * b0 - b2) * cosa + 3 * b1 * cos2a + 2 * b2 * cos3a)
	cap = (cap1 + cap2 + cap3 + cap4 + cap5 + cap6 + cap7 + cap8) / 288

	return sectors - triangle + cap
}

solve(1.11073213677147211458454234765)
solve(1.11073213677147211458454234766)
solve(1.11073213677147211458454234767)
quit
\end{lstlisting}

\newpage
\section*{Output of \texttt{ bc -l smooth.bc}}
\lstset{basicstyle=\footnotesize\ttfamily, keywordstyle=\ttfamily,
	showstringspaces=false, tabsize=2 }
\begin{lstlisting}

a  1.11073213677147211458454234765
b0 -.310039083801076651082339282471989123313522134773361340881533
b1  .882420100742466054972684950196239802835532559877981625163414
b2  .134980967580652222210035504352072567832852870051500822113567
.555360368646626116048170223491013283449047332557401932142645

a  1.11073213677147211458454234766
b0 -.310039083801076651082339283665828267575871461698968187486903
b1  .882420100742466054972684951978550527811372929700478056761308
b2  .134980967580652222210035503671000058497596453386962662777430
.555360368646626116048170223491013283449047332557401932142238

a  1.11073213677147211458454234767
b0 -.310039083801076651082339284859667411838220788624575033935232
b1  .882420100742466054972684953760861252787213299522974488131076
b2  .134980967580652222210035502989927549162340036722424503516382
.555360368646626116048170223491013283449047332557401932142553
\end{lstlisting}


\begin{thebibliography}{9}

	\bibitem{ABC+06}
		H.~Alt, K.~Buchin, O.~Cheong, F.~Hurtado, C.~Knauer, A.~Schulz, and S.~Whitesides.
		Small boxes for carpenter's rules.
		Manuscript, 2006.

	\bibitem{CD05}
		G.~C\u{a}linescu and A.~Dumitrescu.
		The carpenter's ruler folding problem.
		In J.~Goodman, J.~Pach, and E.~Welzl, editors,
		\emph{Combinatorial and Computational Geometry},
		Mathematical Sciences Research Institute Publications,
		volume 52, pages 155--166, 2005.

	\bibitem{CD15}
		K.~Chen and A.~Dumitrescu.
		Nonconvex cases for carpenter's rulers.
		\emph{Theoretical Computer Science},
		586:12--25, 2015.

	\bibitem{KL07}
		O.~Klein and T.~Lenz.
		Carpenter's rule packings --- a lower bound.
		In \emph{Abstracts of 23rd European Workshop on Computational Geometry},
		pages 34--37, 2007.


\end{thebibliography}
\end{document}